\documentclass[aip,twocolumn,apl,graphicx]{revtex4}
\usepackage{graphicx}
\usepackage{dcolumn}
\usepackage{bm}

\begin{document}

\title{Few-hundred GHz Carbon Nanotube NEMS} 

\author{J.~O. Island, V. Tayari, A.~C. McRae, A.~R. Champagne}
\email[]{a.champagne@concordia.ca}
\affiliation{Department of Physics, Concordia University, Montreal, Quebec, H4B 1R6 Canada}

\date{\today}

\begin{abstract}
We study 23 to 30 nm long suspended single-wall carbon nanotube quantum dots and observe both their stretching and bending vibrational modes. We use low-temperature DC electron transport to excite and measure the tubes' bending mode by making use of a positive feedback mechanism between their vibrations and the tunneling electrons. In these nano-electro-mechanical-systems (NEMS), we measure fundamental bending frequencies $f_{bend}\approx$ 75 - 280 GHz, and extract quality factors $Q \sim 10^{6}$. The NEMS' frequencies can be tuned by a factor of two with tension induced by mechanical breakjunctions actuated by an electrostatic force, or tension from bent suspended electrodes.
\end{abstract}


\maketitle 

Single-wall carbon nanotubes (SWCNTs) offer the prospect of making ultra-high frequency, $f$, and quality factor, $Q$, nano-electromechanical systems (NEMS) due to their small mass, large bulk modulus and tunable electro-mechanical properties \cite{Poncharal99, Sazonova04}. There is a considerable ongoing effort to understand the physics of electron-vibron ($e-v$) coupling for the various vibronic modes of carbon nanotubes \cite{LeRoy04,Vitali04, Witkamp06,Garcia-Sanchez07, Mariani09, Huttel09, Leturcq09, Steele09, Lassagne09, Cavaliere10, Schmid12}. An understanding of how the strength of these couplings can be controlled with mechanical strain \cite{Mariani09} could lead to the development of NEMS whose vibrational quantum state could be manipulated to create mechanical qubits \cite{Poot11}. The SWCNT bending mode is of particular interest for potential applications, since its frequency, $f_{bend}$, can be tuned by adjusting the length and tension of the nanotube, and is very sensitive to the tube's environment (adsorbed particles, forces). In high-$Q$ devices, the bending mode vibrations can couple strongly to electron transport \cite{Usmani07,Lassagne09,Steele09, Schmid12}, and be used to develop high-resolution mass sensors \cite{Jensen08, Chiu08, Chaste12} and high-$f$ oscillators \cite{Peng06,Laird11}. Developing SWCNT-NEMS whose fundamental mode $f_{bend}$ is beyond the few-GHz range remains a major challenge both in terms of fabrication, as devices must be a few 10s of nm long, and detection, since measuring frequencies in the 10s or 100s of GHz can be extremely challenging.

We report the observation of strain-tunable bending mode resonances up to $\approx$ 280 GHz, and $Q$-factors of the order of $10^{6}$ in 23 to 30 nm long suspended SWCNT QDs. The $e-v$ coupling for a bending vibron, $\lambda_{bend}$, is much smaller than the ones for the stretching, $\lambda_{stretch}$, or breathing, $\lambda_{breath}$, vibrons \cite{Mariani09}. However, the effective coupling between the tunneling current and bending mode is enhanced in high-$Q$ devices, where a large population of bending vibrons can be created due to a positive feedback between the electron flow and the vibrations \cite{Usmani07, Steele09, Lassagne09, Schmid12}. We use DC electron transport in the quantum dot (QD) regime, $T = 4.2$ K unless specified, to excite both bending and stretching vibrons, and measure their frequencies through their effects on conductivity. We demonstrate that $f_{bend}$ can be tuned via strain applied to the tube, and extract $e-v$ coupling factors $\lambda_{stretch} \sim1$, $\lambda_{bend}\sim 10^{-3}$. These results demonstrate that SWCNTs can be used as extremely high frequency NEMS, up to a few 100s of GHz, and that their tunability makes them good candidates to study the quantum mechanics of macroscopic systems.

\begin{figure}
\includegraphics [width=3.25in]{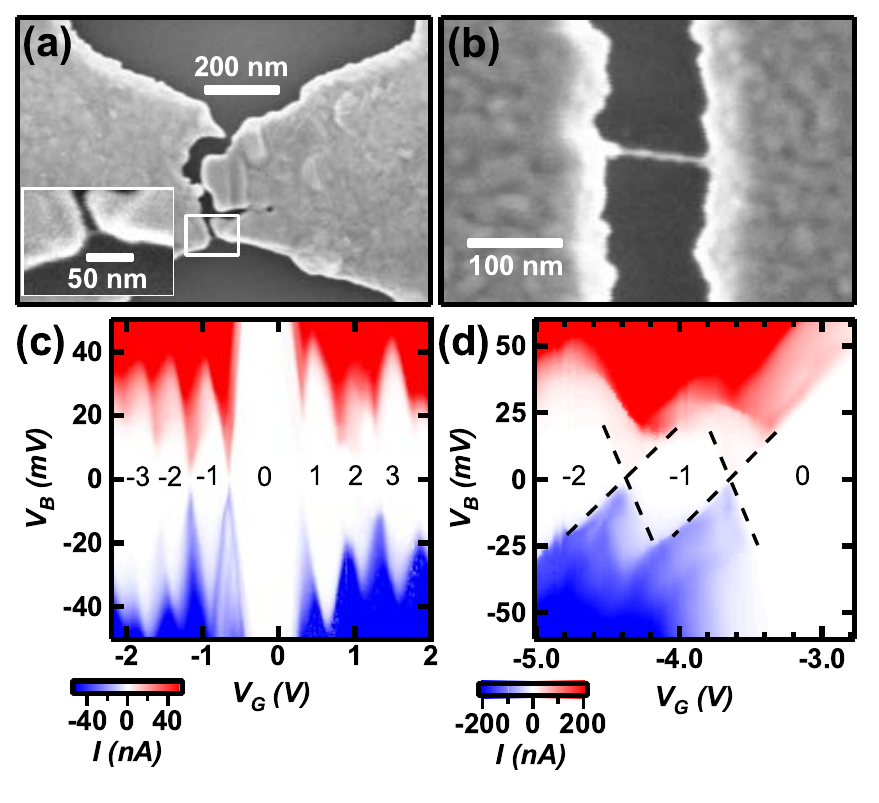}
\caption{\label{}(Color online.) Suspended SWCNT quantum dots. (a) Top-view SEM image showing an electromigrated gold breakjunction across which a 22 nm long suspended SWCNT (Device A) is clearly visible (inset). (b) Top view SEM image showing a 78 nm long suspended SWCNT. A gold meniscus from the left contact extends over the tube and shortens the length of the suspended quantum dot (Device C). (c) and (d) Two-dimensional $I-V_{B}-V_{G}$ Coulomb blockade data for Devices A and C respectively. The positive (negative) numbers in the Coulomb diamonds refer to the number of electrons (holes) in the QD ground state.}
\end{figure}

 We fabricated several ultra-short, 10 to 80 nm, suspended SWCNT devices, see Fig.\ 1a-b and the Supporting Information (SI). We observed stretching vibron modes in the five devices that we studied in detail, and bending mode resonances in three of these devices (Devices A, B, C). We focus our discussion on the later three devices, while data for the other two devices, $\approx$ 10 nm long, are shown in section 1 of the SI. In Fig. 1a and its inset, we see scanning electron microscope (SEM) images of the central portion of bow-tie suspended gold electrodes (breakjunction) contacting a 22 $\pm$ 4 nm long section of SWCNT (Device A). Figure 1b, shows Device C, where a 78 nm long section of suspended SWCNT is contacted by rectangular gold electrodes. The complete sample fabrication procedure for these devices is detailed elsewhere \cite{Island11}, and briefly summarized below.

To fabricate bow-tie breakjunctions on top of SWCNTs (Devices A and B in main text, and Devices D and E in SI), we start with heavily-doped Si wafers with a 300 nm-thick SiO$_{2}$ film on their top side. The Si substrate acts as a back-gate electrode. We grow SWCNTs by chemical vapor deposition \cite{Jin07}, and use AFM to measure their diameter to be $d$ = 2 $\pm$ 0.6 nm. We use e-beam lithography to define 40-nm thick gold bow-tie junctions (no adhesion layer) on top of selected SWCNTs. The bow-ties are approximately one micron long and 350 nm wide at their center, Fig.\ 1a and Figs. S2-S3, and connected via 3 $\mu$m-wide gold wires. We use a wet buffered oxide etch (BOE) to remove SiO$_{2}$ from under the gold bow-tie junctions to suspend them. The final fabrication step is to create a gap in the suspended gold bridges by electromigration \cite{Park99,Pasupathy05, Champagne05,Ward11, Island11} to uncover nm-sized sections of the SWCNTs. To do so, we ramp up a bias voltage, $V_{B}$, across the gold bridges and electromigrate our devices while they are immersed in liquid Helium or in high-vacuum ($\leq 10^{-6}$ Torr) at $T \approx 4.2 $K. We carefully control the rate of electromigration with a custom feedback software to adjust the size of the gap and avoid damaging the SWCNTs \cite{Island11}. The inset of Fig.\ 1a shows an enlargement of the lower portion of the breakjunction of Device A, where the short section of SWCNT across the gap is visible. This position matches the location of the SWCNT before gold deposition. To fabricate Device C, shown in Fig.\ 1b, we directly exposed two rectangular gold electrodes on top of the tube using e-beam lithography, and then suspended the tube with a BOE etch. Figure 1b shows contrast variations along the tube which suggests that the gold meniscus from the left contact extends over much of the suspended tube, shortening the freely suspended section of Device C.

Device A is a semiconducting tube, and enters the QD regime at low-temperature due to Schottky barriers at the Au/SWCNT interfaces. Figure 1c shows the low-temperature current, $I$, versus $V_{B}$, and gate voltage, $V_{G}$, data for Device A. We observe a clear single-electron transistor effect with on and off states, defining a Coulomb diamond for each charge ground state of the QD \cite{vonDelft01}. The diamonds are labeled with positive (negative) integers, $N$, corresponding to the number of electrons (holes) on the dot. The wide blockaded region around $V_{G}=0$ corresponds to the charge neutrality point where no carrier is present on the tube \cite{Biercuk08}. The location of this depletion region near zero gate voltage indicates that there is little chemical doping, and that most of the contamination adsorbed on the tube during fabrication was ashen during the electromigration procedure. From the width of the diamonds for $N\neq0$, we extract a gate capacitance $C_{G}=e/\Delta V_{G}$ = 0.31 aF \cite{vonDelft01}. The capacitance per unit length of the tube can be estimated using a wire over a plane model, $C/L_{G} = 2\pi\epsilon/(\cosh^{-1}(h/r))$, where $h$ is the wire to plane distance and $r$ the wire radius. We model $C_{G}$ as two such capacitors in series (SI section 3), respectively with vacuum and SiO$_{2}$ dielectrics. For Device A, $t_{oxide} =$ 100 nm, measured by ellipsometry, and $t_{vac}=$ 35 $\pm$ 5 nm as measured by AFM. We extract $L_{G}=$ 27 $\pm$ 4 nm. This is in agreement with the length $L =$ 22 $\pm$ 4 nm measured by SEM in Fig.\ 1a, and confirms that the capacitance model gives a reasonable estimate of the length. Since $L_{G}\approx L$, a single QD occupies the full length of the exposed SWCNT. Device B was fabricated following the same procedure as Device A, but its SWCNT is metallic and enters the QD regime at low temperature due to imperfect contacts \cite{Biercuk08}. Its Coulomb blockade data, Fig.\ 2b and SI section 4, show that a single QD, whose length is $L_{G}=$ 30 $\pm$ 4 nm, occupies the suspended tube. The $I-V_{B}-V_{G}$ data for Device C in Fig.\ 1d, taken at $T$ = 2.2 K, are also consistent with a single QD, showing only one set of positive and negative slope tunneling thresholds. From the Coulomb diamonds, we extract $L_{G}$ = 23 $\pm$ 4 nm for Device C, indicating that the QD is shorter than the distance between the gold contacts visible in Fig.\ 1b.

\begin{figure}
\includegraphics [width=3.25in]{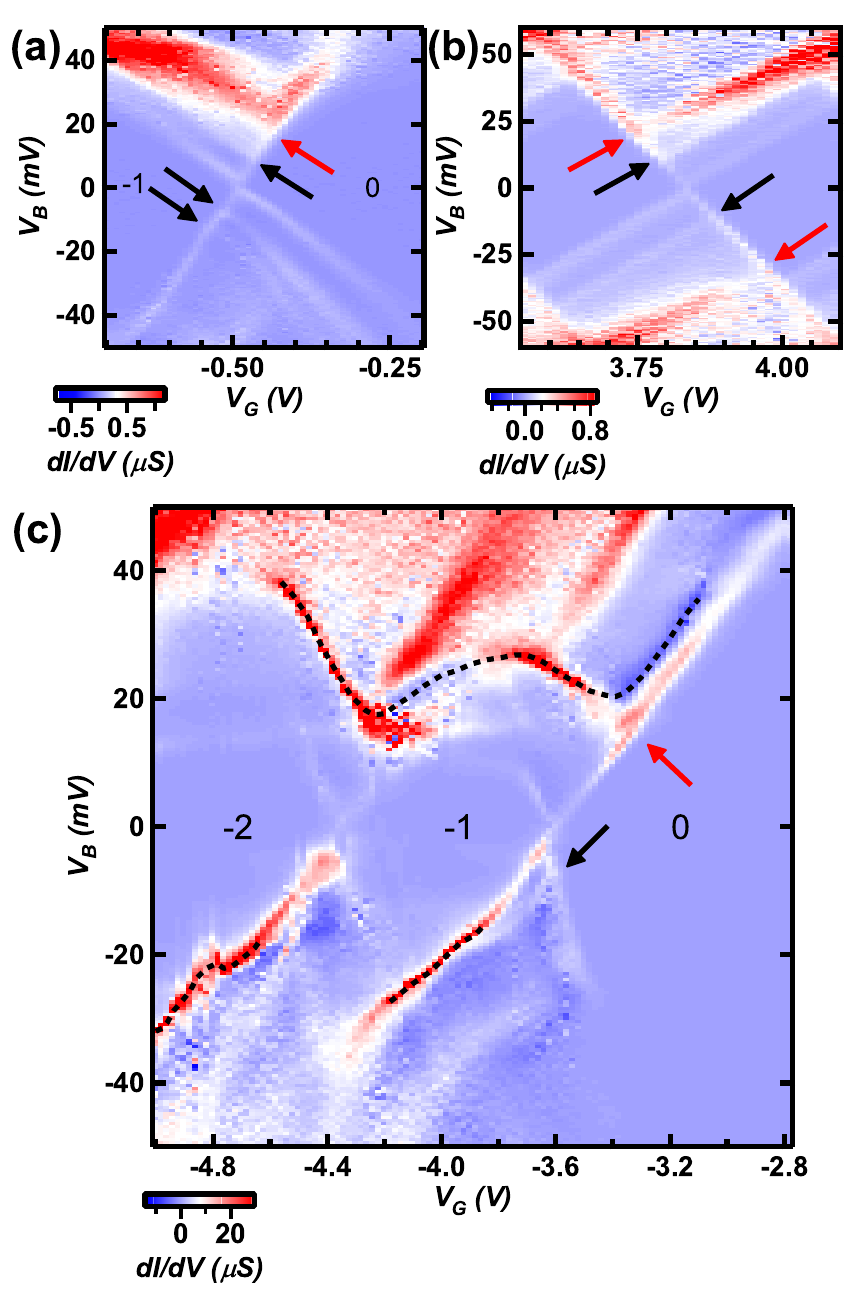}
\caption{\label{} (Color online.) Stretching vibrons. (a), (b), (c) $dI/dV_{B}-V_{B}-V_{G}$ data for Devices A, B and C respectively, before current annealing. The arrows identify the position of some of the vibron sidebands. The red arrows point to sidebands which are not always parallel to the Coulomb thresholds and suggest the emergence of a bending mode resonance. In panel (c), the black dashed lines highlight well resolved bending mode resonances. }
\end{figure}

Figure 2 shows $dI/dV_{B}-V_{B}-V_{G}$ electron spectroscopy data for Devices A, B and C, allowing us to measure the energy of their excited states appearing as resonances running parallel to the Coulomb threshold peaks \cite{vonDelft01}. We extract the energy of these excitations as $E =$ e$V_{B}$ where the tunneling resonance intersects the Coulomb diamonds as indicated by the arrows in Fig.\ 2a-c. The lowest excitation energies are 9, 10 and 7 meV for Devices A, B, and C respectively. These energies are more than an order of magnitude too small to correspond to electronic excitations of 23-30 nm long QDs \cite{Biercuk08}. Rather, they match the energy scale of the longitudinal stretching vibrons \cite{Sapmaz06,Leturcq09}, $\Delta E_{s} = \hbar v_{s} \Delta q$, where $\Delta q = \pi/L_{s}$, $L_{s}$ is the length of the oscillator, and $v_{s} =\sqrt{Y/\rho}$ is the average group velocity \cite{Dresselhaus00}. Our devices have tube diameters $d\approx$ 2 nm, mass density $\rho=$ 870 kg/m$^3$, and Young's modulus $Y\approx$ 1 TPa, giving $v_{s}= 3.4\times 10^4$ m/s. The extracted length of the longitudinal vibrons in the 3 devices are respectively $L_{s}$ = 8, 7 and 10 nm. These lengths are of the same scale, but shorter than the size of the QDs, similarly to what has been reported previously in suspended SWCNT devices where the discrepancy was attributed to localized vibrons \cite{Leturcq09,Cavaliere10,Fabio10}.

The excitation energies of the first vibronic sideband, extracted at each Coulomb diamond (charge state) between $V_{G}$ = -2 and 7 V for Device B show no significant change (Fig. S4). $V_{G}$ creates an electrostatic force which bends the bow-tie suspended gold contacts and can strain the suspended tube by up to a few percent (SI section 2). The absence of strain dependence for the excitations is consistent with stretching vibrons, as their energy is unaffected by strains up to several percent\cite{Wang09}, as opposed to bending mode vibrations whose energies are strongly strain dependent\cite{Sapmaz03}. Bending mode vibrons couple quadratically to electrons and have a much smaller $e-v$ coupling than stretching vibrons \cite{Mariani09}. For 23-30 nm long SWCNTs, bending vibrons have an energy scale of $\sim$ 0.1 meV which is below our temperature limited energy resolution. The radial breathing mode of our SWCNTs has a comparable energy scale to the one for stretching vibrons, $\Delta E_{b} =$ 28 meV/$d$ where $d$ is the nanotube's diameter. However, the breathing mode is an optical mode and its coupling to electrons is expected to be an order of magnitude smaller than for the stretching mode in our devices\cite{Mariani09}. The breathing mode has only been observed in transport experiments when the electrons tunnel in the SWCNTs with their momentum perpendicular to the tube's axis, as in STM experiments \cite{Vitali04,LeRoy04}.

Therefore, we expect that the stretching vibrons couple the most strongly to electron transport in our devices, and we ascribe the excited state sidebands in Fig.\ 2 to stretching vibrons. For an electron tunneling on a QD while simultaneously creating stretching vibrons, the tunneling probability is proportional to the square of the overlap of the QD vibronic wavefunction before (ground state) and after (excited state) tunneling, and $dI/dV\propto e^{-g}g^{n}/n!$, where $g =\lambda_{stretch}^{2}$ and $n$ is the number of vibrons generated during the tunneling process ($n = $ 1 is the first sideband, and so on)\cite{Koch05, Leturcq09}. Hence, the observation of multiple stretching sidebands with similar amplitudes, as in Fig. 2a-b, is only possible for large $e-v$ coupling, when $\lambda_{stretch} \approx$ 1. For $\lambda_{stretch}>$ 1, the ground state to ground state transition, $n = 0$, is exponentially suppressed at low $V_{B}$ where the energies of the electrons are not sufficient to create vibrons (Franck-Condon blockade) \cite{Sapmaz06,Leturcq09,Cavaliere10,Fabio10}. We observe this effect in Fig.\ 2c for Device C, where $dI/dV$ along the Coulomb threshold is heavily suppressed near $V_{B}=$ 0. Close inspection of the positive Coulomb threshold line for the $N = 0$ charge state, shows that its intensity steps up around $V_{B}=$ +7 mV indicating that the $n = 1$ sideband of the stretching mode is present, but suppressed by the FC blockade, while the sideband for $n=2$ around $V_{B}=$ +15 mV is much stronger. Further evidence of strong $e-v_{stretch}$ coupling, and high $Q$-factor, is given by the nearly horizontal co-tunneling resonance \cite{Huttel08} which originates where the stretching sideband meets the Coulomb threshold at $V_{B}=$ +15 mV. We conclude that $\lambda_{stretch}$ is larger than 1 in Device C, and comparable to 1 in A and B. A small difference in the tubes' diameters may explain this variation since $\lambda_{stretch}\propto 1/\sqrt{d}$ \cite{Mariani09}.

Figure 2c shows additional features in $dI/dV$ (dashed black lines) which are not parallel to the Coulomb threshold signaling the bending mode resonance of the NEMS. The coupling of electrons to bending vibrons, $\lambda_{bend}\approx 40\times L[\mu m]/(\pi d[nm])^{3}$, for the fundamental flexural mode is small \cite{Mariani09}. Thus, we would not expect to observe excited state sidebands in $dI/dV$ when $eV_{B}=\hbar\omega_{bend}$ as for the stretching mode, even if our temperature were below the bending vibron energy. However, feedback mechanisms can enhance the effective coupling between bending vibrations and tunneling electrons in QDs. Bending vibrations produce a time dependent gate capacitance, or equivalently an effectively oscillating AC gate voltage. In the QD regime, this AC gate voltage can create AC modulated tunneling rates for the electrons as it moves the energy of the dot across one of the Coulomb blockade tunneling threshold. In turn, the electrons tunneling on/off the QD create an AC electrostatic force on the tube which can further drive the bending mode.  When the current flow through the QD is large enough so that the tunneling rate of electrons coupling to the bending vibrons, $\Gamma$, matches or exceeds $f_{bend}$, and the $Q$-factor is sufficiently high for the vibrons to be long-lived compared to the tunneling rate, this can lead to a positive feedback between the bending mode and tunneling electrons which spontaneously drives large-amplitude bending vibrations \cite{Usmani07}.

From Fig.\ 2c at $V_{G}=$-3.9 V and $V_{B}= -15.2$ mV, where the resonance is sharp and extends slightly into the normally blockaded region, we extract $I =$ 45 $\pm$ 10 nA and $f_{bend}\approx I/e\approx$ 280 GHz. We can compare this frequency with the expected frequency for a tube under zero tension, $T=0$,
\begin{equation}\label{}
f_{bend,T=0}=22.38r/(4\pi L^{2})\sqrt{Y/\rho}
\end{equation}
where $r$ is the tube radius. Using Eq. 1 for Device C with $L = L_{G} = 23$ nm, we find $f_{bend,T=0}=$ 113 GHz. The discrepancy between the measured and expected $f_{bend}$ indicates that the tube is under considerable tension. The predicted dependence of $f_{bend}$ for a SWCNT under high tension is \cite{Sapmaz03}
\begin{equation}\label{}
    f_{bend}=\frac{1}{2\pi}\left[\frac{1.77}{Lr}\sqrt{\frac{T}{\rho}}+\frac{\pi r}{L^{2}}\sqrt{\frac{Y}{\rho}} \right]
\end{equation}
where $T$ is the tension. The tension in Device C was built-in when the device was suspended (BOE etch), and the electrodes contacting the SWCNT bent toward the substrate, Fig. S5. Using tilted SEM and AFM to image the suspended gold electrodes contacting the tube, we confirm that the tube is under a considerable tensile strain, and measure a strain of 9.8$\%$ (SI section 5). This strain level is realistic in SWCNTs which can sustain a strain up to at least 13.7 $\%$ without any plastic deformation \cite{Chang10}. Using the measured strain and Eq. 2, the relationship $f_{bend}\approx I/e$  predicts $I$ = 36 $\pm$ 1 nA (225 GHz), where the uncertainty comes from the uncertainty on the tube's diameter. This expected $f_{bend}$ agrees with the measured current at the resonance, $I$ = 45 $\pm$ 10 nA (281 $\pm$ 63 GHz), where the uncertainty is due to the sharpness of the current step at the resonance. This agreement, together with the detailed measurements on Device A discussed below, confirms the relationship $f_{bend}\approx I/e$  at the bending resonance. We only observed this approximate equality between $f_{bend}$  and the current when the resonance extends inside the normally Coulomb blockaded region. We extract the $e-v_{bend}$ coupling strength in Device C from \cite{Blanter04} $\lambda_{bend} = (F^{2})/(\hbar m_{eff}(2\pi f_{bend})^3)\sim$ 10$^{-3}$, where $F$ is the electrostatic force on the tube, and $m_{eff} = $0.725$ m_{tube}$ is the effective mass of the oscillator.

We note that the bending resonance in Fig.\ 2c emerges from the second stretching mode sideband at positive $V_{B}$ and from the first stretching sideband at negative $V_{B}$. This can be explained by the sudden increase of $I$ at the vibron sidebands which can trigger the onset of the positive feedback necessary for observing the flexural mode. The presence of the stetching vibrons also leads to an increase in the population of bending vibrons because the main decay mode of a stretching vibron is into two bending vibrons \cite{DeMartino09}.

\begin{figure}
\includegraphics [width=3.25in]{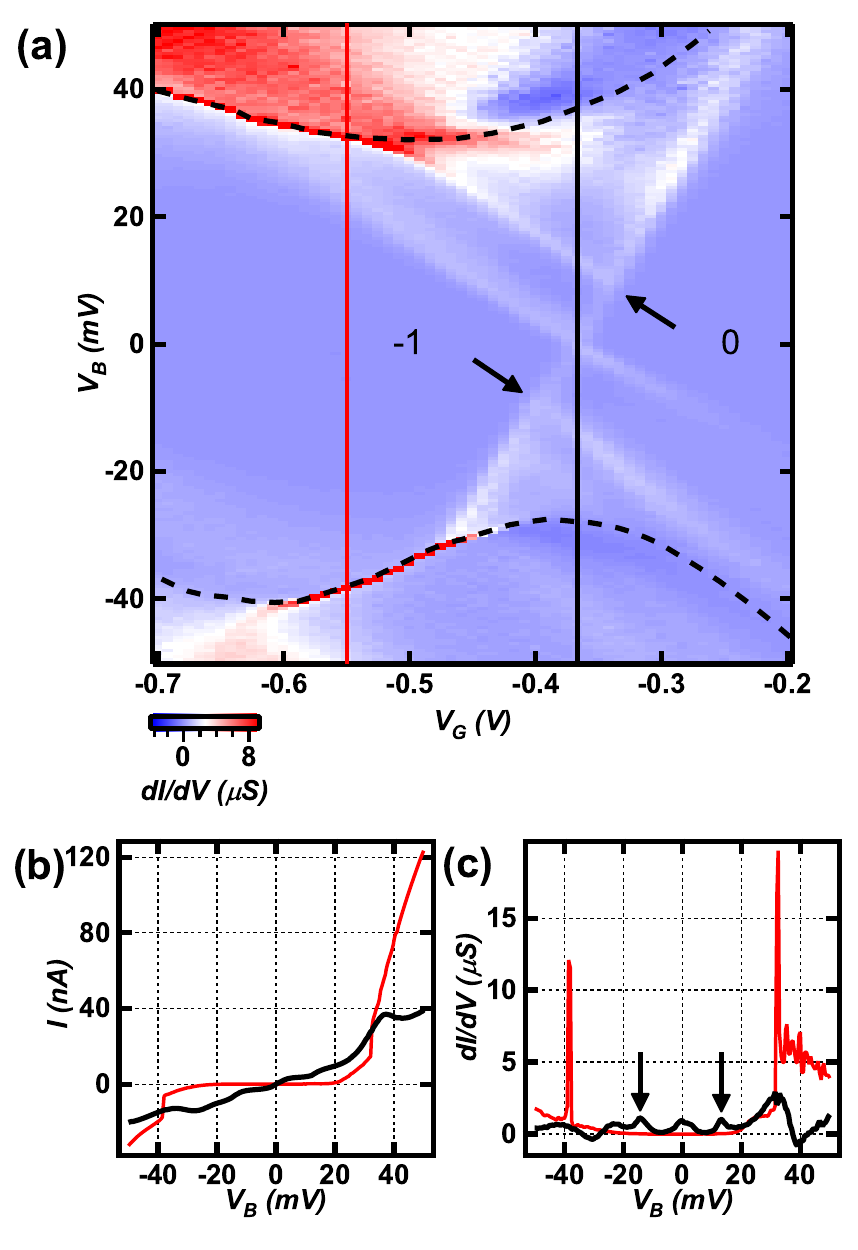}
\caption{\label{} (Color online.) Bending resonances. (a) $dI/dV_{B}-V_{B}-V_{G}$ data for Device A after current annealing. A strong non-linearity in $dI/dV$ appears at high $V_{B}$ (dashed black lines) and signals the excitation of the bending mode. Note that these new features are not parallel to the Coulomb tunneling thresholds and even extend into the normally blockaded region. (b) and (c) One-dimensional $I-V_{B}$ and $dI/dV_{B}-V_{B}$ data cuts respectively, extracted from the red (thin) and black (thick) vertical cuts in (a).}
\end{figure}

We do not observe a sharp bending mode resonance for Devices A and B in Fig.\ 2, but we note broad and strong sidebands at high $V_{B}$ (red arrows) which are not parallel to the Coulomb diamonds. These features suggest that the bending mode non-linearity is washed out due to a low $Q$-factor, which would explain the very gradual rise of $I$ in this region. The current inside the broad resonances at the top left and top right of Figs.\ 2a and 2b respectively are approximately 15 and 10 nA, which correspond to $f_{bend} \sim$ 100 GHz in agreement with the expected $f_{bend,T=0}$ for Devices A and B. We used \textit{in situ} high-current annealing to attempt to remove any remaining surface contamination from Devices A and B which may have been limiting their $Q$. Figure 3a shows the same $dI/dV-V_{B}-V_{G}$ region as Fig.\ 2a for Device A after an annealing step where we flowed $I \approx$ 4 $\mu A$ at $V_{B} =$ 0.33 V for 10 minutes. A very strong bending mode resonance, marked by the black dashed line, is now visible at positive $V_{B}$ where previously in Fig.\ 2a the broad $dI/dV$ feature was present, and an additional resonance is visible at negative $V_{B}$. For Device B, high current annealing at $I \approx$ 4.2 $\mu$A and $V_{B} =$ 0.66 V for 10 minutes did not substantially modify the $dI/dV$ data. The differing effectiveness of the annealing procedure on samples A and B may be due to the higher resistance of Device B. Since the two devices have almost equal lengths, and the tube in Device B is metallic, the higher resistance in B is likely a result of a higher contact resistance. This would mean that a larger portion of the Joule heating power is dissipated in the contacts, thus the temperature rise of Device B during annealing is smaller than in A and the annealing less effective in removing contamination.

There are additional features allowing us to distinguish the stretching mode sidebands from the bending mode $dI/dV$ resonances in Fig.\ 3. In panel (a), as was observed in Device C, the flexural resonance for Device A extends inside what would normally be the Coulomb blockaded region. The observed shape of the bending resonances in Figs.\ 3a and 2c are consistent with both theoretical calculations \cite{Usmani07} and previous experimental observations in devices with lower $f_{bend}$ \cite{Steele09, Lassagne09}. Panels 3b-c show respectively 1-dimensional $I-V_{B}$ and $dI/dV-V_{B}$ cuts along the red (thin) and black (thick) vertical lines in panel (a). The width of the bending mode $dI/dV$ resonance is not temperature limited, and can be much sharper than both the charge tunneling thresholds and the stretching mode side bands. Similarly, the amplitude of the change in $I$ due to the bending resonance can be much bigger than the one due to the longitudinal mode, as seen in the red data in Fig. 3b-c. In Fig. 3a, the bending mode resonances show up as positive $dI/dV$ (PDC) when they have negative slopes for $V_{B} > 0$, or positive slopes for $V_{B} < 0$. Conversely, $dI/dV$ is negative (NDC) when the resonances have positive slopes for $V_{B} > 0$ or negative slopes for $V_{B} < 0$. The same alternating pattern of $(dI/dV)_{bend}$  between PDC and NDC is observed in Fig. 2c for Device C, and in Fig. 4a for Device A over a broader range of $V_{G}$. While the details of this pattern are not currently understood, a possible explanation may relate to the QD energy level alignment with the Fermi level of the electrodes at the Coulomb thresholds that are closest to the bending resonances. A PDC bend mode is observed when the resonance is near the Coulomb thresholds which add an electron to the QD, while a NDC is seen near the transitions which remove an electron from the dot. Thus, the bending vibrons modulate the tunneling rate on (off) the QD at PDC (NDC). Since vibrons are created by electrons tunneling on the QD, at PDC peaks the vibrons are in phase with the incoming electrons and can help scatter them across the dot. On the other hand, at the NDC peaks the vibrons modulate the rate of electrons tunneling off the dot which may cause an out-of-phase driving of the vibrons and lead to more backscattering of the electrons.

The 1-dimensional vertical red (thin) cut in Fig. 3a intersects the bending resonance at negative $V_{B}$ in a region which would normally be inside the Coulomb diamond, and where the relation $f_{bend}\approx \Gamma \approx I/e$ holds. Using a Lorentzian fit, we extract $I =$ 12 $\pm$ 6 nA at the $dI/dV$ resonance for the red (thin) data in Fig. 3b-c located at $V_{B}$ = -38.3 mV and $V_{G}=$ -0.55 V, giving $f_{bend}\approx$ 75 $\pm$ 30 GHz. Using Eq. 1, we calculate for Device A, $f_{bend,T=0} = $ 84 GHz. This close agreement suggests that Device A is not under a significant tension at $V_{G}=$ -0.55 V. The expected strain from the electrostatic bending of the suspended bow-tie cantilevers due to $V_G$ = -0.55 V is negligible, and SEM and AFM imaging of Device A show no apparent static (built-in) bending of the electrodes which would strain the SWCNT (SI section 2). To calculate the $Q$-factor of Devices A and C based on their transport characteristics, we follow \textit{Lassagne et al.} \cite{Lassagne09,Blanter04} who derived for similar samples,
\begin{equation}\label{}
\frac{1}{Q}=\frac{2\pi f_{bend}}{k}\left(\frac{2C_{G}'V_{G}}{\Gamma C_{\Sigma}}\right)^{2} G
\end{equation}
where $C_{G}'$ is the derivative of $C_{G}$ with respect to the vertical displacement of the tube, $k$ is the effective spring constant of the tube, $C_{\Sigma}$ is the total capacitance of the QD, and $G$ the conductance (SI section 6). Using data from Figs.\ 3a and 2c for the bending resonances located inside the normally Coulomb blockaded region, we calculate for Device A, at $V_{G}=$-0.55 V, and Device C, at $V_{G}=$ -3.9 V, respectively $Q=$ 2.9 $\times 10^{6}$ and $1.3\times 10^{6}$. These high $Q_{bend}$ values are expected\cite{DeMartino09} since $Q_{bend}\propto 1/L{^2}$ , and to the best of our knowledge, our devices are the shortest SWCNT-NEMS reported thus far. The calculated $Q$-factors are also consistent with previous work where bending mode resonances were observed in DC transport only when $Q \geq 10^4$ \cite{Steele09, Schmid12}.

\begin{figure}
\includegraphics [width=3.25in]{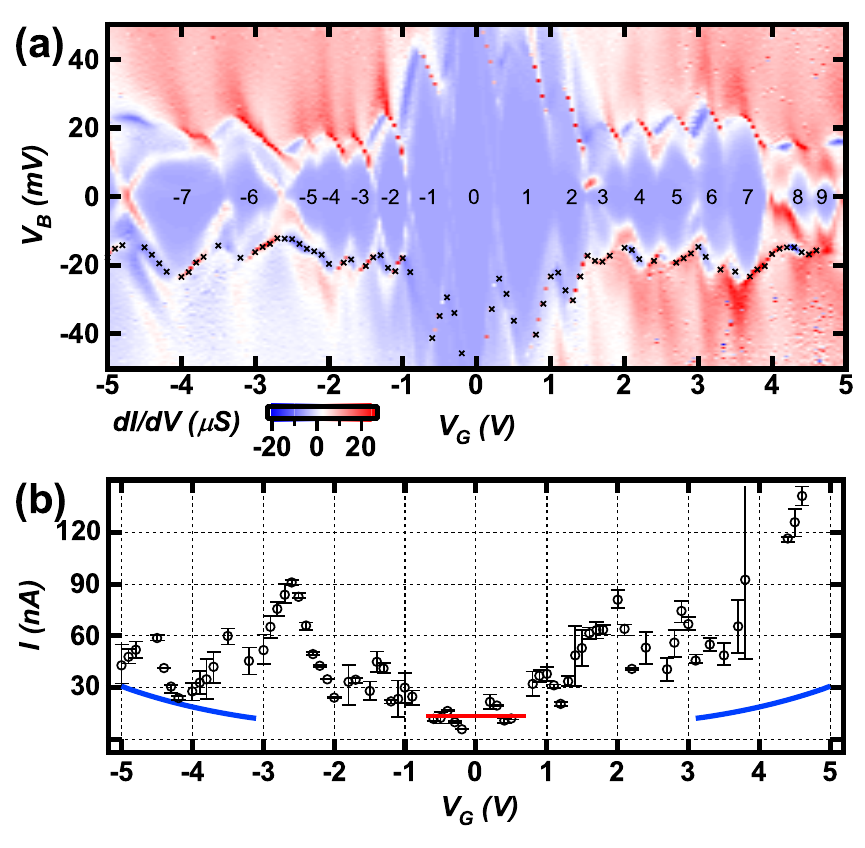}
\caption{\label{} (Color online.) Tension-tuning of $f_{bend}$. (a) $dI/dV_{B}-V_{B}-V_{G}$ data for Device A, after current annealing, over a broad range of $V_{G}$. The gate voltage bends the breakjunction contacts and strains the suspended SWCNT, which tunes $f_{bend}$. The black markers indicate the locations where we extracted the bending mode resonance $I$ values shown in (b). (b) $I$ measured at the bending resonances with $V_{B} < 0$ versus $V_{G}$. The red (thin) and blue (thick) lines show the low and high-tension regime theoretical calculation for $I = ef_{bend}$ valid at the minima of well-resolved Coulomb valleys. Due to the coarseness of the $V_{G}$ data steps, the calculated current is only expected to agree quantitatively with the data at the best resolved (widest) Coulomb valleys located around $V_{G}$ = 0 and -4.2 V.}
\end{figure}

Figure 4 shows data for Device A over a broad range of $V_{G}$. The black markers indicate the locations of the bending $dI/dV$ resonances where we extract the $I$ values shown in Fig. 4b. The positive bias $dI/dV$ bending resonances are left unmarked to show the alternating PDC/NDC pattern at each Coulomb diamond. As $V_{G}$ increases, the electrostatic force between the gate electrode and the suspended bow-tie gold electrodes increases $\propto V_{G}^{2}$. Using the geometry of the two suspended cantilever contacts making up the bow-tie junction we can calculate the vertical displacement of the suspended source and drain electrodes, $\Delta y$, and the longitudinal strain $\Delta L/L$, induced by the moving contacts (SI section 2). This strain applies a tension to the tube which increases rapidly with $V_{G}$, and modifies $f_{bend}$ as given by Eq. 2.

Fig.\ 4b shows $I$ extracted from the negative $V_{B}$ resonance in panel (a) versus $V_{G}$. The current oscillates with the same period as the Coulomb diamonds. The data in Fig. 4a has coarser steps in $V_{G}$ than in Figs.\ 2 and 3, which limits the resolution of the $I$ oscillations in Fig 4b. The effect of the width (resolution) of the Coulomb diamonds on the depth of the current minima is made clear when comparing the data from the far left and far right of Fig. 4a-b. While the high positive gate diamonds are narrow and their $I$ oscillations poorly resolved, the negative $V_{G}$ data show wider diamonds and deeper current minima. The Coulomb valleys around $V_{G} =$ 0 V and -4 V are wide and show the best defined current minima, where we expect $I\approx ef_{bend}$. The red (thin) line and blue (thick) line in Fig.\ 4b are respectively the calculated low-tension regime (essentially Eq.\ 1) and high-tension regime (Eq.\ 2) current expected when the bending resonance is inside the Coulomb blockaded region. These calculations do not include any fitting parameters, and are based on the measured dimensions of Device A (SI section 2). The agreement between the data and theoretical estimate around $V_{G} =$ 0 V and -4 V, combined with a similar agreement for Device C, confirms that we can measure $f_{bend}$ with a reasonable accuracy in DC transport experiments. The frequency around $V_{G} = $ 0 is $\approx$ 80 GHz, and increases up to 150 GHz around $V_{G}$ = -4.2 V where the tube's strain extracted using Eq.\ 2 is 6.0 $\%$.

In summary, using DC electron transport in 23-30 nm long suspended SWCNT QDs we observed both their stretching and bending vibrational modes, and $e-v$ couplings, $\lambda_{stretch} \sim$ 1 and $\lambda_{bend} \sim $10$^{-3}$. The bending mode $dI/dV$ resonance is due to a positive feedback mechanism between electron tunneling and the bending vibrations which turns on when the tunneling rate matches or exceeds $f_{bend}$. Using current annealing, we removed contamination absorbed on one of the oscillators and greatly enhanced the $dI/dV$ resonance associated with the bending mode. We measured fundamental bending mode frequencies up to $\approx$ 280 GHz, roughly two orders of magnitude higher than previously reported fundamental frequencies \cite{Sazonova04, Peng06, Steele09, Lassagne09}, and almost an order of magnitude higher than the highest $f_{bend}$ previously reported \cite{Laird11}. The calculated $Q$-factors of our NEMS are $\sim$10$^{6}$. The flexural frequency $f_{bend}$ can be tuned by a factor of 2 with tension from an electrostatically-actuated mechanical breakjunction, or by a built-in tension due to bent suspended contacts. These extremely high $f_{bend}$ and $Q$-factor NEMS are candidates to create extremely sensitive force and mass sensors, and to explore the quantum mechanics of macroscopic systems \cite{Poot11}. We plan on probing the ultimate limits of SWCNT-NEMS by building even shorter suspended QDs, $\sim$ 5-10 nm \cite{Island11}, with independently tunable charge states and mechanical strain using a gated-mechanical breakjunction approach \cite{Champagne05}.

We thank Serap Yigen and James Porter for technical help. This work was supported by NSERC (Canada), CFI (Canada), FQRNT (Quebec) and Concordia University. We acknowledge using the QNI (Quebec Nano Infrastructure) cleanroom network. Supporting Information: we include data for two additional devices (Devices D and E), detailed imaging of Devices A and C, additional data for Device B, and theoretical calculations of the frequency and quality factors of the NEMS. This material is available free of charge via the Internet at http://pubs.acs.org.

\bibliography{GHz_NEMS}

\end{document}